\documentclass[12pt,preprint]{aastex}
\usepackage{graphics,lscape,ccaption}
\usepackage{amsmath,bm}
\usepackage{times}
\usepackage{epstopdf}
\begin{document}
\slugcomment{Submitted to ApJ}

\title{The Distance Measurement of NGC\,1313 with Cepheids}

\author{Gao Qing\altaffilmark{1}, Wei Wang\altaffilmark{1}, Ji-Feng Liu\altaffilmark{1}}

\altaffiltext{1}{Key Laboratory of Optical Astronomy, National Astronomical Observatories,
Chinese Academy of Sciences, Beijing 100012, China}

\begin{abstract}

We present the detection of Cepheids in the barred spiral galaxy NGC\,1313,  using 
the Wide Field and Planetary Camera 2(WFPC2) on {\it Hubble Space Telescope}. Twenty 
$B$(F450W) and $V$(F555W) epochs of observations spanning over three weeks were obtained, 
on which the profile-fitting photometry of all stars in the monitored field was performed using 
the package HSTphot. A sample of 26 variable stars have been identified to be Cepheids, 
with periods between 3 and 15 days. Based on the derived
period-luminosity (PL) relations in $B$- and $V$-bands, we \textbf{obtain} an extinction-corrected distance modulus of $\bm{\mu_{\rm NGC1313}= 28.32\pm0.10\,{\rm (random)}\pm0.06\,{\rm (systematic)}}$, employing respectively the Large Magellanic Cloud (LMC) as distance zero point calibrators. 
The above moduli correspond to a distance of \textbf{4.61$\bm{\pm0.21}$\,(random)$\bm{\pm0.13}$\,(systematic)\,Mpc}, consistent with previous measurements reported in the literature within
uncertainties. In addition, the reddening to NGC\,1313 is found to be small.
\end{abstract}


\section{INTRODUCTION}
Cepheids are luminous variable stars with \textbf{radial} pulsations, which have been proven to be reliable distance 
indicators with the well-established period-luminosity relation in both observational and theoretical aspects. 
Given their intrinsic visual brightness range from $-2$ to $-6$\,mag, Cepheids are ideal standard candles at 
Galactic and extragalactic distances, and provide useful calibrations for secondary extragalactic distance
indicators including for example Tully-Fisher relation and planetary nebular luminosity functions. 

Thanks to the improved resolution of the \textit{Hubble Space Telescope (HST)}, many Cepheids have been 
discovered in the galaxies in the \textbf{Local Supercluster} and the distances to these galaxies have 
therefore been determined. This campaign extends the distance scale that can be determined reliably by a factor of 10 as compared to the pre-\textit{HST} era, when Cepheids were mostly detected and studied from the ground. For 
a more detailed understanding of Cepheids distance scale and its implications to the local universe, the readers 
are referred to the classic review by \citet{Madore1991}.

The nearby starburst galaxy NGC\,1313 is a spiral galaxy with morphological type SB(s) \citep{deVaucouleurs1991}. 
It is often described as a galaxy in transition between the Magellanic type galaxies and normal disk galaxies such as M33,
given its high rotation velocity and very little or no radial metallicity gradient~\citep{Molla1999}.
Being very active in star formation, NGC\,1313 and its super shell nebula have been extensively studied (e.g. \citealt{Ryder1995, Pellerin2007}), as well as the X-ray sources NGC1313 X-1 and X-2 \citep{Feng2005, Feng2006, Liu2007, Liu2012}
and type{\sc ii} supernova SN1978K~\citep{Ryder1993}. The distance to this galaxy has been determined previously using 
the methods including for example supernova radio flux density and H~{\sc ii} region size distribution, 
which are listed in Table~\ref{distance_estimate} and show modest diversity.
Securing an independent distance measurement using Cepheids in NGC\,1313 is necessary in this context.

\section{Observation and data analysis}

The sky region around NGC\,1313\,X-2 was monitored 20 times over three weeks in 2008 with {\textit HST} WFPC2 in the F450W and 
F555W filters for the program GO-11227 (PI: Jifeng Liu), which was aimed to search and identify periodicity of NGC\,1313\,X-2 in the optical.
For each observation, two 500\,s exposures in F450W were taken, followed by two shots
of 400\,s and 700\,s in F555W. The observations were calibrated with the best calibration files on 2008 August 26,
and standard photometric analysis were performed on each frame using the photometry package HSTphot \citep{Dolphin2000}, which is designed to handle the undersampled point-spread functions (PSF) in WFPC2 images.
The package conducts procedures for cleaning cosmic rays, removing hot pixels, fitting PSF, and applying aperture corrections to secure precise photometry in the VEGAmag photometric system. 

In total, 32482 sources are detected and identified as well-photometered stars in the 5.7 arcmin$^2$ of NGC\,1313 covered 
by the WFPC2 field-of-view. Further investigations will be based on them.

\subsection{Variable Star Search and Cepheids Identification}

Only well-photometered point sources are included for the search of variable stars. Following a similar approach as
\cite{Dolphin2003} in their study of variables in Sextans, we define a variable star with 
magnitude variation $\sigma_{\rm mag} \geqslant0.15$~mag and chi square $\chi_{\nu}^2\geqslant2$,
where  
$$
\sigma_{mag} \equiv ( \frac{1}{40} (\sum_{i=1}^{20}{(B_{i}-\overline{B})^{2}}   +  \sum_{i=1}^{20}{(V_{i}-\overline{V})^{2}}))^{2} 
$$
and 

$$
\chi_{\nu}^2 \equiv (\frac{1}{40}(\sum_{i=1}^{20}{\frac{(B_{i}-\overline{B})^{2}}{\sigma_{i,B}^2}} + \sum_{i=1}^{20}{\frac{(V_i - \overline{V})^2}{\sigma_{i,V}^2}}))^{\frac{1}{2}}
$$

Periods of these variable star candidates are determined using a modified Lafler-Kinman algorithm \citep{Lafler1965}, but with a relatively loose criterion of minimum $\Theta\leqslant 0.85$ in either $B$ or $V$ light curves for in order not to miss any periodic variables. Among the total 32, 482 stars, 1616 are found to be variables, and $\sim65$\% of them show periodic variations.

\begin{figure}[htbp]
\centering
\includegraphics[clip=true,bb=50 50 265 268,width=10cm]{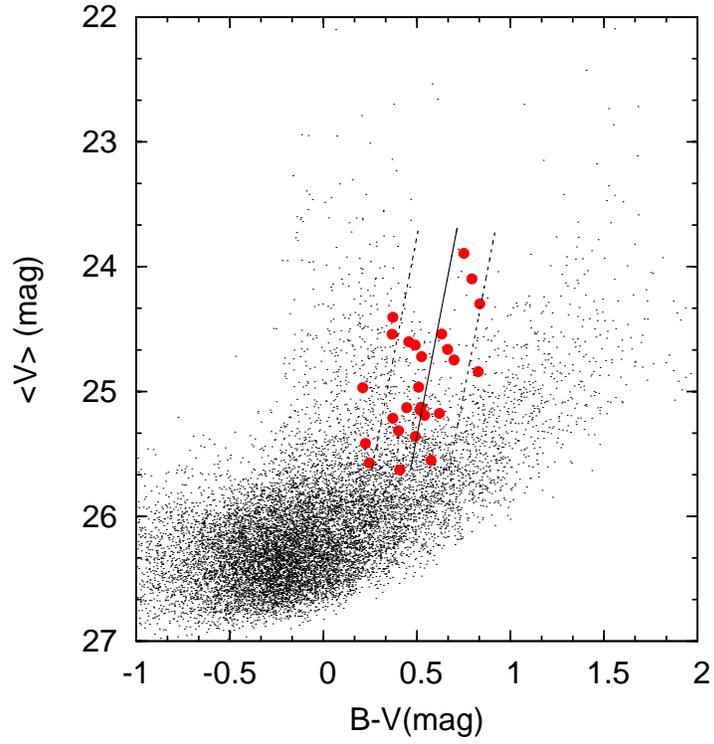}
\caption{\label{CMD_fig} \bf Color-magnitude diagram for the HST/WFPC2 field under current study. Black dots
represent all the detected stars, while the red filled circles represent the 26 Cepheids identified in this paper.
The dashed lines represent the zero reddening instabillity strip of LMC Cepheids and its 2 $\bm{\sigma}$ width.}
\end{figure}

The light curves of these variable stars with periodic variations were visually examined, and 26 of them are found to be Cepheid on the basis of their distinctively rapid brightening, followed by a long, linear decay phase. They are presented in the color-magnitude diagram (CMD) as in Fig.~\ref{CMD_fig} as red
filled circles, along with all 32, 482 stars in black dots.

\subsection{Light curves and average magnitudes for Cepheids}

In Table 2 and 3, $V$- and $B$-bands time-series photometry for the Cepheids are presented for each epoch, 
along with photometric uncertainties, respectively.
\textbf{
We determined the pulsation period and the average magnitudes in $\bm{V}$- and $\bm{B}$- bands for each Cepheid by using template fitting procedures.
The template fitting method based on Fourier analysis and Principal component analysis (PCA) was firstly 
introduced by \cite{Stetson1996} and \cite{Kaluzny1998}.
It can obtained more robust periods and precisely measure average magnitudes even in few observational epochs and bad-sampled light curves.
\cite{Yoachim2009} and \cite{Pejcha2012} expanded on these methods to build new templates for analysis of HST Cepheids data.
The templates of \cite{Yoachim2009} include $\bm{BVI}$ bands and short-period Cepheids ($\bm{<}$ 10 days) in the analysis.
It was ideal for characterizing our discovered Cepheids.}

\textbf{
The training set included the light curves of fundamental (FU) type Cepheids stars comes from OGLE database for the 
LMC \citep{Udalski1999a}, SMC \citep{Udalski1999b} and VizieR database for the Milky Way \citep{Ochsenbein2000}.
We put in a single fake $\bm{I}$- band observation as there are no $\bm{I}$ band photometry.
Considering there have fewer data points in $\bm{B}$- band and deblending problems can cause outlier points in light curves, 
we fixed all additional PCA parameters in fitting procedure.
As a result, the magnitudes and periods were determined through template fitting.
The best-fit light curve templates also generated from obtained parameters.}

\textbf{
The light curves of photometry data along with their best-fit templates (solid lines) 
in $\bm{V}$- and $\bm{B}$- are plotted in Figure~2 with their Ids, periods, $\bm{V}$ and $\bm{B}$ magnitudes.
The average $\bm{V}$- and $\bm{B}$- magnitudes and their measurement uncertainties for 
each Cepheid are given in Table 4.
The $\bm{\chi^2}$/dof for the fitting are given in the last column.}

\section{PL relations and the distance to NGC\,1313}

\textbf{
The phase-weighted apparent $\bm{V}$ and $\bm{B}$ PL relations for the Cepheids in NGC\,1313 are plotted in Figures 3.
The filled circles show the 24 Cepheids used for the distance calculating, while the 2 filled diamonds show the excluded ones, namely Cepheid\,10 and 11.
The NGC\,1313 data are superimposed on the PL relation for LMC Cepheids (open circles) from updated OGLE II catalog \citep{Udalski1999a}.
The solid lines are fits to the LMC data.}

To determine the distance to the galaxy NGC\,1313,
the apparent PL relations need to be calibrated with the absolute PL relations in LMC galaxies, 
which contain a large number of Cepheids and have a well determined distance modulus of 
$\bm{18.493\pm0.008\,{\rm (statistical)}\pm0.047\,{\rm (systematic)}}$
\textbf{
\citep{Pietrzyski2013}.}
\textbf{
From the OGLE II LMC Cepheid catalog \citep{Udalski1999a}, we use their revised PL relations which using 650 
FU Cepheids in $\bm{V}$- band and 662 FU Cpeheids in $\bm{B}$- band :}
\begin{equation}
\begin{matrix}
\bm{M_V({\rm LMC}) = (-2.779 \pm 0.031)\,{\rm log}\,P + (17.066 \pm 0.021)}     \\
\bm{M_B({\rm LMC}) = (-2.439 \pm 0.046)\,{\rm log}\,P + (17.368 \pm 0.031) }
\end{matrix}
\end{equation}
with \textbf{standard deviation of 0.160\,mag at $\bm{V}$ and 0.239\,mag at $\bm{B}$}, respectively.
\textbf{Here $\bm{M_V({\rm LMC})}$ and $\bm{M_B({\rm LMC})}$ are extinction-corrected values using the reddening of E($\bm{B-V}$)=0.147 to LMC.}

In order to avoid incompleteness bias in fitting a slope to the NCG\,1313 data,
only the intercept of the linear regression was fitted, while keeping the slope fixed to the LMC values.
We shift the NGC\,1313 magnitudes to match those of LMC, and determine the relative distance moduli by minimizing the
best-fit chi square of the linear fitting of the combined data set.
As a result, the apparent relative distance moduli between NGC\,1313 and the LMC are measured to be
$\bm{\Delta \mu_{V} = 9.893 \pm 0.049}$ mag and $\bm{\Delta \mu_{B}= 9.872 \pm 0.057}$ mag.

\subsection{The true distance modulus and its uncertainties}
 
The above derived distance moduli need to be corrected for foreground extinction and reddening effect. 
\textbf{The true absolute wavelength-independent distance modulus to NGC\,1313 can be expressed as a function of
the relative apparent modulus $\bm{\Delta\mu}$ between NGC\,1313 and LMC, 
the LMC distance modulus, and the extinction for NGC\,1313, as:}
\begin{equation}
\label{muV}
\bm{
\mu_{V} = \Delta \mu_V + \mu^{\rm LMC} = \mu + A_{V} 
}
\end{equation}
\begin{equation}
\label{muB}
\bm{
\mu_{B} = \Delta \mu_B + \mu^{\rm LMC} = \mu + A_{B} 
}
\end{equation}
\begin{equation}
\label{mu}
\bm{
\mu = \mu_V - R_V(\mu_{B} - \mu_{V})
}
\end{equation}

Where the reddening law of $R_V = A_{V}/E(B-V)=3.3$ for NGC\,1313 and the Magellanic Clouds~\citep{Ferrarese1996}.
The true modulus to \textbf{$\bm{\mu^{LMC} = 18.493}$\,mag\,\citep{Pietrzyski2013}.
By subtracting Equation~\ref{muB} to Equation~\ref{muV}, we yield the extinction of NGC\,1313:}
\begin{equation}
\label{ebmv}
\bm{E(B-V)=A_{B}-A_{V}=\Delta \mu_B-\Delta \mu_V}
\end{equation}
\textbf{
Based on Equation~\ref{muB} and Equation~\ref{muV}, we found the apparent distance moduli for NGC\,1313
are then $\bm{\mu_V = 28.386 \pm 0.049}$\,mag and $\bm{\mu_B = 28.365 \pm 0.057}$\,mag.
Those two modulus agree to each other within errors.
That implied zero reddening.
We computed an error-weighted average value as the true distance modulus to NGC\,1313
$\bm{\mu = 28.377 \pm 0.038}$\,mag instead of applying dereddening precedure of Equation~\ref{mu}.
}

In the above results,errors for the distance modulus consist of two independent factors: random errors and systematic errors.
The random errors are calculated using Equation~(18) of Saha et al. (2000), propagated mainly from the errors on stellar photometry and intrinsic widths of the PL relation in $V$ and $B$:
\begin{equation}
\bm{\sigma^2_{random} =  (\sigma^2_V + \sigma^2_B) + (\sigma_{\rm PL, B}^2+\sigma_{\rm PL, V}^2)}
\end{equation}

The first item in the right side is the estimated photometric errors for the mean $V$ and $B$ magnitudes, 
as propagated through the dereddening procedure, where $\bm{\sigma_V = 0.029}$ and $\bm{\sigma_B = 0.070}$\,mag 
respectively. The second item is associated with the intrinsic width of the PL relation (i.e., a given Cepheid 
may not be on the mean ridge line of the PL relation), which is $\bm{\sigma_{\rm PL, V}=0.046}$\,mag, and 
$\bm{\sigma_{\rm PL, B}=0.050}$\,mag.

Several systematic errors also come into play: uncertainties due to the distance modulus of zero point galaxy, which
is $\bm{0.008\,{\rm (statistical)}\pm0.047\,{\rm (systematic)}}$\,mag~for LMC~\citep{Pietrzyski2013}, errors derived from the observed scatter in PL relation of LMC and (all lower than 0.02 mag), uncertainties due the metallicity dependency of the PL relation. 
\cite{Walsh1997} performed multifiber emission-line spectrophometry of 33 H~{\sc ii} regions in NGC\,1313,
and derived a flat disc metallicity of ${\rm log}_{10}{\rm (O/H)} = -3.6\pm 0.1$, which is very similar to that for LMC with
\textbf{$\bm{{\rm log}_{10}{\rm (O/H)} = -3.5\pm 0.08}$ \citep{Ferrarese2000}}.
\textbf{On the observational side, in the joint fits of LMC, M81, and NGC\,4258, \cite{Gerke2011} found a luminosity dependence on metallicity of 
$\bm{\gamma_{\mu} = -0.62^{+0.31}_{-0.35}}$ mag dex$\bm{^{-1}}$. 
Applying this correction, our Cepheid distance moduli are corrected by:
$\bm{\Delta \mu_{met} = (0.62 \pm 0.35)([O/H]_{NGC1313} - 8.5) = -0.062 \pm 0.035}$.
On the theoretical side, \cite{Bono2010} used the pulsation models with nebular oxygen abundance $\bm{7.56 \leq 12+log(O/H) \leq 8.88}$.
It shows the metallicity effect on $\bm{\Delta \mu_0 (WBV)}$ is quite strong and they found $\bm{\gamma(WBV) = -0.58 \pm 0.03 }$ mag dex$\bm{^{-1}}$.
Applying this correction, our Cepheid distance moduli are corrected by:
$\bm{\Delta \mu_{met} = (0.58 \pm 0.03)([O/H]_{NGC1313} - 8.5) = -0.058 \pm 0.003}$.
}

\textbf{
We list all our known sources of error in table 5. The distance modulus based on our 24 Cepheids is $\bm{28.32\pm0.10\,}$(random)$\bm{\pm0.06}$\,(systematic),
giving a corresponding distance of $\bm{4.61\pm0.21}$\,(random)$\bm{\pm0.13}$\,(systematic)\,Mpc.
}

\section{Discussion}

We have derived a Cepheid distance to NGC\,1313 based on $V$- and $B$-bands
{\textit HST} WFPC2 observations.  A total of 26 Cepheids have been found with
periods ranging from \textbf{3.12 to 14 days}, with their phase-weighted average
magnitudes in both bands and periods in $V$ determined.  Apparent $V$ and $B$
distance moduli to NGC\,1313 have been derived, employing the LMC
Cepheids samples to be zero point reference galaxies. After applying extinction
corrections, the true distance modulus to NGC\,1313 is
\textbf{$\bm{28.32\pm0.10\,}$(random)$\bm{\pm0.06}$\,(systematic)}
using LMC for the
calibration of distance. These moduli correspond to distance of
\textbf{$\bm{4.61\pm0.21}$\,(random)$\bm{\pm0.13}$\,(systematic)\,Mpc.}
 The Cepheid distance found in this
work is slightly larger than previous measurements, but are consistent within
errors with most results, especially those obtained using tip of RGB stars
(summarized in Table 1).

An important factor on distance measurement is that Cepheids are oscillating in
two modes --- fundamental (FU) and first over-tone (FO).  Cepheids with periods
shorter than 8 days are likely to be oscillating as a first over-tone pulsator
that appears brighter and bluer than fundamental Cepheids.  To confuse this two
type Cepheids would make the distance underestimated.  Since the PL relations
are calculated by only using FU Cepheids from OGLE database, we must make sure
that our 26 Cepheids are all belong to this type. \cite{Beaulieu1995} studied
large sample of Cepheids light-curve in 490 nm in LMC, and found that the first
over-tone (FO) Cepheids have a strong tendency to be more symmetric, and having
lower amplitudes in light curves than the FU Cepheids.
Fig.~\ref{Cepheid_classification} presents the amplitudes against their periods
of 26 Cepheids in $V$ and 6 in $B$ with detections in all 20 observations,
along with the 94 LMC Cepheids from \cite{Beaulieu1995}.  It is obvious that
their LMC FU  and FO Cepheids occupy quite different regions in the
Amplitude-Period (AP) plane. All of our Cepheids are within the area where LMC
FU Cepheids are located, and they should be of FU type.

The extinction of NGC\,1313 is also obtained when deriving the distance modulus.
Based on \textbf{Equation~\ref{ebmv}}, we found the extinction to NGC 1313 is $\bm{E(B-V) = -0.01\pm 0.07}$.
The value for the estimated extinction is low compare with previous studies
around the region of NGC 1313 X-2.  An extinction of E(B-V) = 0.11 mag can be
obtained from the \citet{Schlegel1998} dust map as the Galactic average
value within $0.6^\circ$ of NGC1313 X-2.  An extinction of E(B-V) = 0.44 mag
(corresponding to $n_H = 2.7 \times  10^{21} cm^{-2}$) is inferred from the
power-law fit to the X-ray spectrum of NGC 1313 X-2~\citep{Miller2003},
which also includes contribution from circumbinary materials.
The extinction value we obtained ($<$ 0.03) through Cepheids suggests that the
average extinction within $2^\prime$ of NGC1313 X-2 is slightly smaller than
the Galactic average value within $0.6^\circ$ of NGC1313 X-2.

\section*{Acknowledgments}

We thank Drs. Yu Bai, Yi Hu Song Wang and Hang Gong for very helpful discussions. {JFL acknowledges the support from National Science Foundation of China through NSFC grants \#11233004 and \#11333004. }{WW is partially supported by NSFC grant \#11203035 and 11233004 for this work.}

\bibliography{ref,new}
\clearpage
\begin{table}
\centering
\caption{\label{distance_estimate}The previous and current determinations of the distance to NGC\,1313 using various 
methods, as listed in Column~2. Only measurements uncertainties (random) are taken into account in the error budgets given in Column~1 except this work.}
\begin{tabular}{l l r}

 \hline
 \hline
  Distance (Mpc)   &    Method           & References      \\
 \hline
 \noalign{\smallskip}
 4.13$\stackrel{\scriptscriptstyle+0.11}{\scriptscriptstyle-0.12}$   & Tip of RGB (TRGB)                & Bryan Mendez et al 2002 \\
 4.15$\stackrel{\scriptscriptstyle+0.06}{\scriptscriptstyle-0.06}$   & TRGB                                  & \cite{Tully2006} \\
 4.27$\stackrel{\scriptscriptstyle+0.06}{\scriptscriptstyle-0.06}$   & TRGB                                  & \cite{Rizzi2007} \\
 4.07$\stackrel{\scriptscriptstyle+0.20}{\scriptscriptstyle-0.21}$   & TRGB                                  & \cite{Grise2008}\\
 4.39$\stackrel{\scriptscriptstyle+0.04 }{\scriptscriptstyle-0.04 }$ & TRGB                                  & \cite{Jacobs2009} \\
 3.7$^{\rm a}$                                                                 & Tully-Fisher correlation            & \cite{Tully1988} \\
 3.0                                                                               & H{\sc ii} region size distribution & \cite{Issa1981}    \\
 3.9$\stackrel{\scriptscriptstyle+0.8 }{\scriptscriptstyle-0.8 }$      & SN 1978K radio flux density & \cite{Weiler1998}\\
 $\bm{4.61\pm0.21\pm0.13}$                                                      & Cepheids, calibrated with LMC   & this work \\
 \hline
\end{tabular}    
\begin{list}{}{}
\item{$^{\rm a}$ Assuming the Hubble constant H$_{0}$=75.0 km\,s$^{-1}$\,Mpc$^{-1}$.}
\end{list}
\end{table}

\clearpage
\begin{table}[htpb]
\centering
\small
\caption{$V$ photometry of the 26 Cepheids discovered in this work.}
\begin{tabular}{l l l l l l l }
\hline
\noalign{\smallskip}
JD& C1 & C2 &C3 &C4 &C5 &C6  \\
\noalign{\smallskip}
\hline
54607.92485 &24.489$\pm$0.073 &24.841$\pm$0.074 &24.352$\pm$0.063 &25.072$\pm$0.101 &24.726$\pm$0.130 &24.450$\pm$0.049       \\ 
54608.05749 &24.276$\pm$0.065 &24.871$\pm$0.074 &24.250$\pm$0.053 &24.957$\pm$0.072 &24.585$\pm$0.108 &24.354$\pm$0.048       \\
54609.0561  &24.007$\pm$0.054 &24.450$\pm$0.054 &24.461$\pm$0.053 &24.649$\pm$0.070 &24.807$\pm$0.092 &24.420$\pm$0.047       \\
54610.05472 &24.566$\pm$0.092 &24.468$\pm$0.055 &24.686$\pm$0.064 &24.517$\pm$0.054 &24.854$\pm$0.091 &23.983$\pm$0.037       \\
54611.05263 &24.579$\pm$0.062 &24.536$\pm$0.059 &24.918$\pm$0.087 &24.719$\pm$0.061 &25.076$\pm$0.108 &23.814$\pm$0.035       \\
54612.11791 &24.728$\pm$0.109 &24.468$\pm$0.077 &25.206$\pm$0.093 &24.663$\pm$0.059 &25.559$\pm$0.180 &23.958$\pm$0.044       \\
54613.24291 &24.874$\pm$0.121 &24.755$\pm$0.086 &25.025$\pm$0.145 &24.857$\pm$0.082 &24.872$\pm$0.089 &24.173$\pm$0.047       \\
54614.24152 &24.577$\pm$0.078 &24.969$\pm$0.103 &24.259$\pm$0.054 &24.806$\pm$0.078 &24.591$\pm$0.071 &24.084$\pm$0.040       \\
54615.24083 &24.092$\pm$0.062 &24.915$\pm$0.091 &24.709$\pm$0.082 &24.931$\pm$0.070 &24.864$\pm$0.101 &24.309$\pm$0.045       \\
54616.11097 &24.468$\pm$0.079 &24.707$\pm$0.066 &24.826$\pm$0.080 &25.067$\pm$0.094 &24.903$\pm$0.103 &24.287$\pm$0.044       \\
54617.16791 &24.600$\pm$0.071 &24.295$\pm$0.050 &24.947$\pm$0.093 &24.734$\pm$0.063 &25.062$\pm$0.166 &24.354$\pm$0.047       \\
54618.7061  &24.787$\pm$0.106 &24.595$\pm$0.062 &25.252$\pm$0.111 &24.394$\pm$0.049 &25.504$\pm$0.161 &24.487$\pm$0.053       \\
54619.23597 &24.877$\pm$0.101 &24.599$\pm$0.062 &24.688$\pm$0.083 &24.583$\pm$0.067 &24.967$\pm$0.096 &24.668$\pm$0.061       \\
54620.16444 &24.974$\pm$0.118 &24.728$\pm$0.066 &24.258$\pm$0.052 &24.602$\pm$0.060 &24.704$\pm$0.098 &24.561$\pm$0.057       \\
54621.23388 &24.375$\pm$0.060 &24.897$\pm$0.080 &24.830$\pm$0.084 &24.834$\pm$0.068 &24.945$\pm$0.112 &24.607$\pm$0.065       \\
54622.16235 &24.266$\pm$0.057 &25.306$\pm$0.141 &24.944$\pm$0.076 &24.803$\pm$0.063 &25.016$\pm$0.118 &24.410$\pm$0.047       \\
54623.16097 &24.361$\pm$0.061 &24.705$\pm$0.070 &25.081$\pm$0.099 &25.017$\pm$0.074 &25.076$\pm$0.113 &24.471$\pm$0.050       \\
54624.08944 &24.743$\pm$0.101 &24.370$\pm$0.050 &25.048$\pm$0.098 &25.210$\pm$0.085 &25.291$\pm$0.163 &23.969$\pm$0.035       \\
54625.15888 &24.857$\pm$0.107 &24.456$\pm$0.059 &24.668$\pm$0.063 &24.590$\pm$0.056 &24.815$\pm$0.105 &23.857$\pm$0.038       \\
54626.62555 &24.827$\pm$0.090 &24.563$\pm$0.059 &24.482$\pm$0.061 &24.451$\pm$0.052 &24.854$\pm$0.105 &24.052$\pm$0.039       \\
\noalign{\smallskip}
\hline
\noalign{\smallskip}
\noalign{\smallskip}
  C7 & C8 &C9 &C10  & C11 &C12 & C13  \\
\noalign{\smallskip}
\hline
\noalign{\smallskip}
 24.566$\pm$0.054 &23.678$\pm$0.044 &23.803$\pm$0.033 &24.760$\pm$0.088  &24.518$\pm$0.063 &24.988$\pm$0.119 &25.464$\pm$0.149 \\    
 24.540$\pm$0.053 &23.601$\pm$0.034 &23.799$\pm$0.043 &24.581$\pm$0.104  &24.582$\pm$0.077 &24.899$\pm$0.089 &25.450$\pm$0.168 \\    
 24.268$\pm$0.044 &23.681$\pm$0.044 &23.640$\pm$0.030 &24.424$\pm$0.065  &24.763$\pm$0.076 &24.641$\pm$0.070 &25.415$\pm$0.128 \\   
 24.202$\pm$0.044 &23.709$\pm$0.046 &23.824$\pm$0.040 &24.373$\pm$0.056  &24.526$\pm$0.073 &24.406$\pm$0.059 &24.751$\pm$0.074 \\    
 24.360$\pm$0.050 &23.839$\pm$0.040 &23.973$\pm$0.041 &24.688$\pm$0.077  &24.370$\pm$0.060 &24.673$\pm$0.077 &25.412$\pm$0.172 \\    
 24.461$\pm$0.052 &23.876$\pm$0.041 &24.084$\pm$0.042 &...               &24.543$\pm$0.098 &24.826$\pm$0.085 &25.177$\pm$0.111 \\ 
 24.484$\pm$0.052 &24.050$\pm$0.047 &24.213$\pm$0.054 &24.344$\pm$0.057  &24.952$\pm$0.126 &24.870$\pm$0.090 &25.543$\pm$0.160 \\    
 24.633$\pm$0.059 &24.131$\pm$0.048 &24.371$\pm$0.049 &24.399$\pm$0.059  &24.495$\pm$0.064 &24.997$\pm$0.098 &25.376$\pm$0.139 \\    
 24.419$\pm$0.049 &24.210$\pm$0.061 &24.458$\pm$0.063 &24.795$\pm$0.095  &24.515$\pm$0.066 &24.108$\pm$0.079 &24.927$\pm$0.090 \\    
 24.160$\pm$0.041 &24.065$\pm$0.048 &24.395$\pm$0.053 &24.701$\pm$0.072  &24.568$\pm$0.070 &24.522$\pm$0.064 &25.078$\pm$0.164 \\    
 24.331$\pm$0.048 &23.745$\pm$0.038 &24.303$\pm$0.060 &24.483$\pm$0.066  &24.694$\pm$0.079 &24.533$\pm$0.073 &25.238$\pm$0.124 \\    
 24.508$\pm$0.055 &23.632$\pm$0.036 &24.030$\pm$0.049 &24.461$\pm$0.065  &24.467$\pm$0.066 &24.784$\pm$0.124 &25.567$\pm$0.188 \\    
 24.481$\pm$0.054 &23.683$\pm$0.037 &23.869$\pm$0.037 &24.757$\pm$0.083  &24.752$\pm$0.084 &25.046$\pm$0.109 &...              \\    
 24.578$\pm$0.054 &23.637$\pm$0.034 &23.747$\pm$0.033 &24.597$\pm$0.067  &24.686$\pm$0.073 &24.464$\pm$0.075 &25.022$\pm$0.109 \\    
 24.443$\pm$0.053 &23.783$\pm$0.037 &23.799$\pm$0.034 &24.306$\pm$0.056  &24.680$\pm$0.073 &24.248$\pm$0.053 &25.020$\pm$0.094 \\    
 24.079$\pm$0.039 &23.843$\pm$0.075 &24.053$\pm$0.052 &24.440$\pm$0.058  &24.521$\pm$0.089 &24.552$\pm$0.066 &25.224$\pm$0.122 \\    
 24.284$\pm$0.045 &23.951$\pm$0.045 &24.087$\pm$0.042 &24.593$\pm$0.069  &24.715$\pm$0.077 &24.700$\pm$0.080 &25.323$\pm$0.143 \\    
 24.308$\pm$0.045 &24.136$\pm$0.048 &24.130$\pm$0.048 &24.760$\pm$0.075  &24.799$\pm$0.080 &24.889$\pm$0.086 &25.360$\pm$0.126 \\    
 24.413$\pm$0.050 &24.171$\pm$0.051 &24.358$\pm$0.049 &24.325$\pm$0.056  &24.495$\pm$0.067 &24.890$\pm$0.105 &25.110$\pm$0.106 \\    
 24.566$\pm$0.061 &24.170$\pm$0.052 &24.409$\pm$0.072 &24.528$\pm$0.067  &24.668$\pm$0.077 &24.227$\pm$0.054 &25.124$\pm$0.135 \\    
\noalign{\smallskip}
\hline
\noalign{\smallskip}
\end{tabular}
\end{table}

\begin{table}[htpb]
\centering
\small 
\contcaption{V photometry of 26 Cepheids (continued)}
\begin{tabular}{lllllll}
\hline
\hline
\noalign{\smallskip}
JD& C14 & C15 & C16 &C17 & C18 & C19  \\
\noalign{\smallskip}
\hline
\noalign{\smallskip}
54607.92485  &25.401$\pm$0.126 &25.411$\pm$0.136 &25.334$\pm$0.109 &25.464$\pm$0.149 &24.967$\pm$0.097 &25.259$\pm$0.122   \\ 
54608.05749  &...              &25.407$\pm$0.122 &25.220$\pm$0.139 &25.450$\pm$0.168 &25.211$\pm$0.105 &25.179$\pm$0.095   \\
54609.0561   &...              &25.223$\pm$0.181 &24.818$\pm$0.072 &25.415$\pm$0.128 &...              &25.216$\pm$0.108   \\
54610.05472  &25.083$\pm$0.102 &24.798$\pm$0.072 &25.177$\pm$0.114 &24.751$\pm$0.074 &25.615$\pm$0.156 &24.917$\pm$0.086   \\
54611.05263  &25.328$\pm$0.126 &25.394$\pm$0.126 &25.306$\pm$0.116 &25.412$\pm$0.172 &25.184$\pm$0.148 &25.158$\pm$0.105   \\
54612.11791  &25.408$\pm$0.133 &25.315$\pm$0.115 &25.367$\pm$0.122 &25.177$\pm$0.111 &25.280$\pm$0.116 &25.370$\pm$0.107   \\
54613.24291  &24.649$\pm$0.079 &24.779$\pm$0.095 &24.927$\pm$0.084 &25.543$\pm$0.160 &25.536$\pm$0.152 &25.302$\pm$0.124   \\
54614.24152  &24.916$\pm$0.084 &25.206$\pm$0.104 &24.904$\pm$0.097 &25.376$\pm$0.139 &25.850$\pm$0.189 &24.943$\pm$0.083   \\
54615.24083  &25.284$\pm$0.121 &25.499$\pm$0.135 &25.173$\pm$0.102 &24.927$\pm$0.090 &24.925$\pm$0.091 &25.074$\pm$0.117   \\
54616.11097  &...              &25.120$\pm$0.094 &25.134$\pm$0.095 &25.078$\pm$0.164 &25.142$\pm$0.099 &25.360$\pm$0.122   \\
54617.16791  &24.782$\pm$0.081 &25.040$\pm$0.096 &25.452$\pm$0.166 &25.238$\pm$0.124 &25.581$\pm$0.162 &25.449$\pm$0.150   \\
54618.7061   &25.207$\pm$0.120 &25.368$\pm$0.131 &24.942$\pm$0.090 &25.567$\pm$0.188 &25.169$\pm$0.112 &25.031$\pm$0.097   \\
54619.23597  &25.493$\pm$0.162 &...              &24.954$\pm$0.094 &...              &24.862$\pm$0.087 &25.080$\pm$0.090   \\
54620.16444  &25.537$\pm$0.146 &24.654$\pm$0.066 &25.295$\pm$0.130 &25.022$\pm$0.109 &25.151$\pm$0.100 &25.255$\pm$0.112   \\
54621.23388  &24.755$\pm$0.074 &25.318$\pm$0.132 &25.375$\pm$0.130 &25.020$\pm$0.094 &25.653$\pm$0.203 &25.514$\pm$0.130   \\
54622.16235  &25.115$\pm$0.100 &25.517$\pm$0.152 &25.271$\pm$0.111 &25.224$\pm$0.122 &...              &24.564$\pm$0.080   \\
54623.16097  &25.270$\pm$0.119 &25.488$\pm$0.154 &24.941$\pm$0.084 &25.323$\pm$0.143 &24.949$\pm$0.089 &25.255$\pm$0.107   \\
54624.08944  &25.395$\pm$0.130 &24.698$\pm$0.067 &24.939$\pm$0.081 &25.360$\pm$0.126 &25.184$\pm$0.129 &25.221$\pm$0.090   \\
54625.15888  &24.918$\pm$0.099 &25.201$\pm$0.106 &25.298$\pm$0.114 &25.110$\pm$0.106 &25.713$\pm$0.173 &25.317$\pm$0.139   \\
54626.62555  &25.303$\pm$0.127 &25.556$\pm$0.147 &25.501$\pm$0.141 &25.124$\pm$0.135 &24.897$\pm$0.086 &24.804$\pm$0.081   \\
\noalign{\smallskip}
\hline
\noalign{\smallskip}

 C20  & C21 & C22 & C23 & C24 & C25 & C26  \\
\noalign{\smallskip}
\hline
\noalign{\smallskip}
25.390$\pm$0.170 &25.141$\pm$0.091 &25.551$\pm$0.113 &25.153$\pm$0.086 &25.315$\pm$0.102 &25.789$\pm$0.151  &25.830$\pm$0.193 \\ 
25.152$\pm$0.117 &25.430$\pm$0.119 &25.386$\pm$0.130 &25.142$\pm$0.080 &25.427$\pm$0.115 &25.824$\pm$0.154  &25.702$\pm$0.142 \\
25.640$\pm$0.177 &25.609$\pm$0.154 &25.983$\pm$0.163 &25.250$\pm$0.093 &25.334$\pm$0.112 &24.959$\pm$0.078  &25.225$\pm$0.087 \\
25.673$\pm$0.171 &25.766$\pm$0.179 &25.740$\pm$0.134 &25.217$\pm$0.084 &25.076$\pm$0.084 &25.194$\pm$0.091  &25.339$\pm$0.102 \\
25.366$\pm$0.151 &25.376$\pm$0.120 &25.381$\pm$0.108 &24.564$\pm$0.055 &25.109$\pm$0.104 &25.570$\pm$0.142  &25.693$\pm$0.176 \\
25.415$\pm$0.217 &25.358$\pm$0.164 &25.614$\pm$0.151 &24.795$\pm$0.065 &25.289$\pm$0.114 &25.586$\pm$0.133  &...              \\
25.537$\pm$0.162 &25.710$\pm$0.163 &25.873$\pm$0.159 &25.089$\pm$0.078 &25.205$\pm$0.096 &25.064$\pm$0.084  &25.115$\pm$0.082 \\
25.187$\pm$0.112 &25.937$\pm$0.191 &25.563$\pm$0.137 &25.457$\pm$0.126 &25.493$\pm$0.122 &25.247$\pm$0.099  &25.667$\pm$0.126 \\
25.688$\pm$0.193 &25.129$\pm$0.095 &25.810$\pm$0.150 &25.448$\pm$0.103 &25.012$\pm$0.091 &25.410$\pm$0.109  &25.482$\pm$0.154 \\
25.873$\pm$0.207 &25.596$\pm$0.137 &26.054$\pm$0.179 &24.520$\pm$0.052 &25.170$\pm$0.092 &...               &25.912$\pm$0.160 \\
25.149$\pm$0.114 &25.656$\pm$0.159 &25.174$\pm$0.094 &24.632$\pm$0.062 &25.412$\pm$0.129 &25.245$\pm$0.138  &25.014$\pm$0.083 \\
25.427$\pm$0.178 &25.791$\pm$0.201 &25.993$\pm$0.189 &25.150$\pm$0.087 &...              &25.087$\pm$0.090  &25.858$\pm$0.155 \\
...              &25.272$\pm$0.115 &25.740$\pm$0.154 &25.270$\pm$0.098 &24.789$\pm$0.072 &25.399$\pm$0.130  &25.835$\pm$0.154 \\
24.827$\pm$0.081 &25.493$\pm$0.126 &25.141$\pm$0.085 &25.466$\pm$0.103 &24.958$\pm$0.078 &25.504$\pm$0.118  &25.953$\pm$0.158 \\
25.554$\pm$0.153 &25.739$\pm$0.155 &25.417$\pm$0.105 &24.601$\pm$0.055 &25.116$\pm$0.087 &25.601$\pm$0.128  &25.288$\pm$0.093 \\
...              &25.861$\pm$0.173 &25.718$\pm$0.133 &24.675$\pm$0.057 &25.185$\pm$0.092 &25.106$\pm$0.093  &25.468$\pm$0.106 \\
24.821$\pm$0.088 &25.140$\pm$0.095 &25.435$\pm$0.109 &25.097$\pm$0.085 &25.523$\pm$0.135 &25.234$\pm$0.096  &25.699$\pm$0.134 \\
25.555$\pm$0.206 &25.511$\pm$0.126 &25.382$\pm$0.105 &25.108$\pm$0.080 &24.834$\pm$0.075 &25.490$\pm$0.113  &26.039$\pm$0.167 \\
25.698$\pm$0.185 &25.812$\pm$0.174 &25.919$\pm$0.166 &25.333$\pm$0.095 &25.238$\pm$0.099 &25.770$\pm$0.157  &25.105$\pm$0.082 \\
25.068$\pm$0.106 &25.655$\pm$0.168 &25.183$\pm$0.104 &24.604$\pm$0.057 &25.429$\pm$0.118 &25.056$\pm$0.086  &25.498$\pm$0.113 \\
\noalign{\smallskip}
\hline
\noalign{\smallskip}
\end{tabular}
\end{table}

\begin{table}[htpb]
\centering
\small
\caption{$B$ photometry of the 26 Cepheids discovered in this work.}
\begin{tabular}{l l l l l l l }
\hline
\noalign{\smallskip}
JD& C1 & C2 &C3 &C4 &C5 &C6  \\
\noalign{\smallskip}
\hline
54607.92485 &...              &...              &24.851$\pm$0.104 &...              &24.818$\pm$0.150 &25.176$\pm$0.118       \\ 
54608.05749 &24.853$\pm$0.166 &...              &24.583$\pm$0.086 &25.830$\pm$0.188 &24.704$\pm$0.133 &25.325$\pm$0.109       \\
54609.0561  &...              &25.159$\pm$0.164 &24.992$\pm$0.120 &...              &...              &25.397$\pm$0.112       \\
54610.05472 &25.382$\pm$0.204 &24.985$\pm$0.130 &25.346$\pm$0.152 &25.021$\pm$0.110 &...              &24.624$\pm$0.066       \\
54611.05263 &...              &25.066$\pm$0.129 &...              &25.267$\pm$0.147 &25.096$\pm$0.208 &24.513$\pm$0.068       \\
54612.11791 &...              &25.422$\pm$0.174 &...              &25.349$\pm$0.125 &...              &24.720$\pm$0.072       \\
54613.24291 &...              &25.554$\pm$0.205 &...              &25.666$\pm$0.168 &...              &24.845$\pm$0.096       \\
54614.24152 &...              &...              &24.728$\pm$0.100 &25.454$\pm$0.213 &24.824$\pm$0.178 &24.740$\pm$0.074       \\
54615.24083 &24.659$\pm$0.109 &...              &24.887$\pm$0.117 &...              &...              &25.004$\pm$0.131       \\
54616.11097 &25.069$\pm$0.212 &25.336$\pm$0.165 &...              &25.728$\pm$0.187 &...              &25.180$\pm$0.097       \\
54617.16791 &25.178$\pm$0.207 &24.964$\pm$0.153 &...              &25.600$\pm$0.165 &...              &25.084$\pm$0.098       \\
54618.7061  &...              &25.299$\pm$0.161 &...              &25.105$\pm$0.106 &...              &25.535$\pm$0.140       \\
54619.23597 &24.934$\pm$0.186 &...              &25.515$\pm$0.189 &25.203$\pm$0.121 &24.809$\pm$0.194 &25.411$\pm$0.142       \\
54620.16444 &...              &...              &24.687$\pm$0.102 &25.365$\pm$0.160 &24.733$\pm$0.151 &25.549$\pm$0.138       \\
54621.23388 &24.709$\pm$0.118 &25.255$\pm$0.202 &25.133$\pm$0.155 &25.212$\pm$0.119 &...              &25.322$\pm$0.134       \\
54622.16235 &24.906$\pm$0.181 &25.439$\pm$0.167 &25.517$\pm$0.176 &...              &...              &25.040$\pm$0.092       \\
54623.16097 &25.281$\pm$0.160 &24.950$\pm$0.157 &...              &25.454$\pm$0.172 &...              &25.761$\pm$0.197       \\
54624.08944 &25.182$\pm$0.175 &25.008$\pm$0.153 &...              &25.836$\pm$0.204 &...              &24.700$\pm$0.080       \\
54625.15888 &...              &25.097$\pm$0.145 &25.406$\pm$0.160 &25.122$\pm$0.157 &...              &24.453$\pm$0.075       \\
54626.62555 &...              &25.278$\pm$0.160 &24.901$\pm$0.115 &25.148$\pm$0.110 &...              &24.741$\pm$0.099       \\
\noalign{\smallskip}
\hline
\noalign{\smallskip}
\noalign{\smallskip}
  C7 & C8 &C9 &C10  & C11 &C12 & C13  \\
\noalign{\smallskip}
\hline
\noalign{\smallskip}
25.095$\pm$0.129 &24.326$\pm$0.085 &24.434$\pm$0.057 &24.997$\pm$0.153  &25.009$\pm$0.131 &25.360$\pm$0.196 &25.422$\pm$0.156 \\
24.764$\pm$0.084 &24.165$\pm$0.066 &24.399$\pm$0.058 &25.101$\pm$0.148  &25.199$\pm$0.172 &25.307$\pm$0.192 &25.506$\pm$0.147 \\
24.384$\pm$0.104 &24.333$\pm$0.101 &24.230$\pm$0.060 &24.701$\pm$0.103  &24.998$\pm$0.129 &25.063$\pm$0.144 &...              \\
24.529$\pm$0.091 &24.329$\pm$0.074 &24.662$\pm$0.069 &24.708$\pm$0.102  &...              &24.807$\pm$0.152 &...              \\
24.813$\pm$0.105 &24.560$\pm$0.091 &24.782$\pm$0.077 &25.017$\pm$0.157  &25.019$\pm$0.141 &25.032$\pm$0.151 &25.991$\pm$0.199 \\
24.713$\pm$0.083 &24.902$\pm$0.129 &25.092$\pm$0.101 &...               &...              &25.359$\pm$0.210 &25.347$\pm$0.131 \\
24.966$\pm$0.115 &24.866$\pm$0.155 &25.034$\pm$0.092 &24.937$\pm$0.172  &...              &...              &25.865$\pm$0.203 \\
24.953$\pm$0.101 &24.947$\pm$0.140 &25.333$\pm$0.147 &24.551$\pm$0.134  &24.987$\pm$0.192 &...              &25.996$\pm$0.203 \\
24.778$\pm$0.087 &24.989$\pm$0.139 &25.243$\pm$0.126 &...               &25.051$\pm$0.163 &24.679$\pm$0.108 &25.834$\pm$0.181 \\
24.401$\pm$0.065 &24.623$\pm$0.142 &25.054$\pm$0.090 &...               &25.305$\pm$0.183 &24.832$\pm$0.154 &25.056$\pm$0.097 \\
24.835$\pm$0.095 &24.547$\pm$0.105 &25.197$\pm$0.110 &24.679$\pm$0.147  &...              &24.792$\pm$0.176 &25.916$\pm$0.204 \\
24.757$\pm$0.085 &24.219$\pm$0.067 &24.868$\pm$0.094 &24.823$\pm$0.160  &25.101$\pm$0.186 &25.221$\pm$0.180 &...              \\
24.927$\pm$0.132 &24.399$\pm$0.096 &24.539$\pm$0.071 &25.074$\pm$0.199  &25.021$\pm$0.149 &...              &...              \\
24.911$\pm$0.097 &24.283$\pm$0.071 &24.219$\pm$0.053 &25.334$\pm$0.193  &...              &24.943$\pm$0.135 &25.108$\pm$0.098 \\
24.875$\pm$0.094 &24.585$\pm$0.090 &24.540$\pm$0.065 &24.869$\pm$0.149  &25.063$\pm$0.141 &24.895$\pm$0.127 &25.633$\pm$0.145 \\
24.435$\pm$0.068 &24.612$\pm$0.148 &24.773$\pm$0.085 &24.731$\pm$0.116  &25.084$\pm$0.145 &...              &...              \\
24.628$\pm$0.078 &24.703$\pm$0.103 &24.838$\pm$0.087 &25.123$\pm$0.195  &...              &25.350$\pm$0.208 &25.780$\pm$0.170 \\
24.685$\pm$0.080 &25.159$\pm$0.160 &24.873$\pm$0.083 &25.218$\pm$0.176  &...              &25.385$\pm$0.214 &24.985$\pm$0.090 \\
25.016$\pm$0.106 &25.033$\pm$0.144 &25.175$\pm$0.106 &24.819$\pm$0.163  &...              &25.160$\pm$0.173 &25.535$\pm$0.143 \\
24.917$\pm$0.098 &24.818$\pm$0.115 &25.370$\pm$0.116 &24.664$\pm$0.100  &24.737$\pm$0.115 &24.753$\pm$0.128 &...              \\
\noalign{\smallskip}
\hline
\noalign{\smallskip}
\end{tabular}
\end{table}

\begin{table}[htpb]
\centering
\small 
\contcaption{$B$ photometry of the 26 Cepheids discovered in this work.}
\begin{tabular}{lllllll}
\hline
\hline
\noalign{\smallskip}
JD& C14 & C15 & C16 &C17 & C18 & C19  \\
\noalign{\smallskip}
\hline
\noalign{\smallskip}
54607.92485   &...                  &...              &25.780$\pm$0.215 &...              &...              &25.798$\pm$0.179 \\ 
54608.05749   &...                  &...              &25.478$\pm$0.167 &...              &25.567$\pm$0.209 &25.859$\pm$0.194 \\
54609.0561    &...                  &25.578$\pm$0.201 &25.456$\pm$0.159 &...              &...              &...              \\
54610.05472   &...                  &25.262$\pm$0.141 &...              &25.269$\pm$0.167 &...              &25.175$\pm$0.107 \\
54611.05263   &...                  &...              &...              &...              &25.357$\pm$0.179 &25.555$\pm$0.154 \\
54612.11791   &...                  &25.561$\pm$0.204 &...              &...              &...              &...              \\
54613.24291   &25.256$\pm$0.210     &25.046$\pm$0.126 &25.373$\pm$0.157 &...              &...              &...              \\
54614.24152   &...                  &...              &25.408$\pm$0.164 &...              &...              &25.398$\pm$0.192 \\
54615.24083   &...                  &...              &...              &25.043$\pm$0.207 &25.168$\pm$0.156 &25.727$\pm$0.195 \\
54616.11097   &...                  &...              &...              &...              &25.448$\pm$0.202 &25.742$\pm$0.178 \\
54617.16791   &...                  &...              &...              &...              &...              &...              \\
54618.7061    &...                  &...              &25.493$\pm$0.189 &...              &...              &...              \\
54619.23597   &...                  &...              &...              &...              &...              &25.594$\pm$0.180 \\
54620.16444   &...                  &25.117$\pm$0.132 &...              &25.127$\pm$0.192 &...              &...              \\
54621.23388   &25.129$\pm$0.143     &25.500$\pm$0.190 &...              &...              &...              &...              \\
54622.16235   &...                  &...              &...              &...              &...              &25.271$\pm$0.118 \\
54623.16097   &...                  &...              &...              &...              &25.146$\pm$0.152 &25.714$\pm$0.177 \\
54624.08944   &...                  &25.186$\pm$0.159 &...              &...              &...              &25.913$\pm$0.212 \\
54625.15888   &25.016$\pm$0.136     &...              &25.650$\pm$0.210 &...              &...              &25.807$\pm$0.203 \\
54626.62555   &...                  &...              &...              &24.986$\pm$0.211 &25.226$\pm$0.151 &25.447$\pm$0.197 \\
\noalign{\smallskip}
\hline
\noalign{\smallskip}

 C20  & C21 & C22 & C23 & C24 & C25 & C26  \\
\noalign{\smallskip}
\hline
\noalign{\smallskip}
 ...               &25.632$\pm$0.188 &...              &25.678$\pm$0.213 &...              &...              &...              \\ 
 ...               &...              &...              &25.708$\pm$0.176 &...              &...              &...              \\
 ...               &...              &...              &25.656$\pm$0.152 &...              &25.437$\pm$0.147 &...              \\
 ...               &...              &...              &25.790$\pm$0.162 &25.611$\pm$0.186 &...              &...              \\
 25.167$\pm$0.167  &25.528$\pm$0.186 &...              &24.915$\pm$0.090 &25.343$\pm$0.148 &...              &...              \\
 ...               &25.467$\pm$0.201 &...              &25.516$\pm$0.134 &...              &...              &...              \\
 ...               &...              &...              &25.714$\pm$0.161 &...              &25.276$\pm$0.169 &25.455$\pm$0.172 \\
 ...               &...              &...              &...              &...              &25.479$\pm$0.163 &...              \\
 ...               &...              &...              &25.720$\pm$0.185 &25.184$\pm$0.134 &...              &...              \\
 ...               &...              &...              &24.931$\pm$0.089 &25.563$\pm$0.182 &...              &...              \\
 ...               &...              &...              &25.253$\pm$0.115 &...              &...              &25.667$\pm$0.178 \\
 ...               &...              &...              &25.901$\pm$0.184 &...              &...              &...              \\
 ...               &...              &...              &25.817$\pm$0.188 &...              &...              &...              \\
 ...               &...              &25.722$\pm$0.166 &25.804$\pm$0.182 &25.512$\pm$0.197 &...              &...              \\
 ...               &...              &25.676$\pm$0.168 &25.053$\pm$0.096 &...              &...              &25.594$\pm$0.165 \\
 ...               &...              &...              &25.053$\pm$0.096 &...              &25.190$\pm$0.124 &...              \\
 25.046$\pm$0.148  &...              &...              &25.426$\pm$0.130 &...              &...              &...              \\
 ...               &25.584$\pm$0.197 &...              &...              &...              &...              &...              \\
 ...               &...              &...              &...              &...              &...              &25.391$\pm$0.138 \\
 25.325$\pm$0.195  &...              &25.554$\pm$0.146 &24.953$\pm$0.089 &...              &25.497$\pm$0.161 &...              \\
\noalign{\smallskip}
\hline
\noalign{\smallskip}
\end{tabular}
\end{table}

\begin{table}
\centering
\caption{\label{Cepheid_info}The final sample of 26 Cepheids, with their IDs, 
coordinates and periods presented in Columns~1-4. Final photometry, photometric 
error, and number of observations in $V$ are given in Columns~5-7, while 
those in $B$ in Columns~8-10. \textbf{The $\bm{\chi^2}$/dof for the fitting are given in the last column.}} 
% In Column~12, the sequence of the CCD chip in which each Cepheid locates are listed.
\begin{tabular}{l l l l c c c c c c c}
\noalign{\smallskip}
\hline
\hline
\noalign{\smallskip}
ID    & R.A.  & Decl.  & P      & $V$      &error &N$_{V}$ &$B$      &error &  N$_{B}$  & $\chi^2$/dof  \\
      & (J2000)     & (J2000) &(days)  &(mag)   & (mag)    &        & (mag) & (mag )&   &                    \\
\noalign{\smallskip}
\hline
 C1 & 3:17:57.12  & $-66$:34:19.2  & \bf6.50  & \bf24.540 & \bf0.029 & 20  & \bf25.172 & \bf0.088 &  10   &   \bf2.6          \\ 
 C2 & 3:18:23.76  & $-66$:34:33.6  & \bf7.39  & \bf24.663 & \bf0.018 & 20  & \bf25.326 & \bf0.051 &  14   &   \bf1.4          \\ 
 C3 & 3:18:23.76  & $-66$:35:49.2  & \bf5.86  & \bf24.721 & \bf0.031 & 20  & \bf25.246 & \bf0.071 &  12   &   \bf3.0          \\ 
 C4 & 3:18:30.96  & $-66$:34:55.2  & \bf8.00  & \bf24.747 & \bf0.020 & 20  & \bf25.446 & \bf0.048 &  16   &   \bf1.7         \\ 
 C5 & 3:18:20.16  & $-66$:34:51.6  & \bf5.85  & \bf24.970 & \bf0.023 & 20  & \bf25.180 & \bf0.076 &  6    &   \bf0.8          \\ 
 C6 & 3:18:28.8   & $-66$:34:26.4  & \bf14.00 & \bf24.297 & \bf0.019 & 20  & \bf25.133 & \bf0.040 &  20   &   \bf3.3          \\ 
 C7 & 3:18:25.92  & $-66$:34:44.4  & \bf6.54  & \bf24.405 & \bf0.021 & 20  & \bf24.777 & \bf0.038 &  20   &   \bf3.4          \\ 
 C8 & 3:18:20.64  & $-66$:35:09.6  & \bf10.50 & \bf23.894 & \bf0.012 & 20  & \bf24.644 & \bf0.027 &  20   &   \bf1.5          \\ 
 C9 & 3:18:26.4   & $-66$:34:04.8  & \bf11.50 & \bf24.098 & \bf0.013 & 20  & \bf24.892 & \bf0.023 & 20    &   \bf1.7          \\ 
C10 & 3:18:18.72  & $-66$:34:40.8  & \bf4.00  & \bf24.541 & \bf0.022 & 19  & \bf24.909 & \bf0.047 &  17   &   \bf1.8         \\ 
C11 & 3:18:18.96  & $-66$:34:44.4  & \bf3.75  & \bf24.601 & \bf0.030 & 20  & \bf25.057 & \bf0.079 &   12  &   \bf3.3         \\ 
C12 & 3:18:18.96  & $-66$:34:51.6  & \bf5.50  & \bf24.629 & \bf0.021 & 20  & \bf25.119 & \bf0.048 &  16   &   \bf1.5         \\ 
C13 & 3:18:26.16  & $-66$:34:12.0  & \bf4.08  & \bf24.841 & \bf0.022 & 20  & \bf25.668 & \bf0.043 & 14    &   \bf1.2         \\ 
C14 & 3:18:22.32  & $-66$:34:58.8  & \bf3.90  & \bf25.150 & \bf0.036 & 17  & \bf25.669 & \bf0.107 &  3    &   \bf0.7         \\ 
C15 & 3:18:25.44  & $-66$:35:02.4  & \bf3.56  & \bf25.130 & \bf0.029 & 19  & \bf25.575 & \bf0.071 & 7     &   \bf1.3          \\ 
C16 & 3:18:26.16  & $-66$:35:06.0  & \bf5.00  & \bf25.175 & \bf0.022 & 20  & \bf25.796 & \bf0.065 & 7     &   \bf0.9          \\ 
C17 & 3:18:22.08  & $-66$:35:06.0  & \bf5.24  & \bf25.215 & \bf0.025 & 19  & \bf25.587 & \bf0.087 &  4    &   \bf0.8          \\ 
C18 & 3:18:22.32  & $-66$:34:44.4  & \bf3.79  & \bf25.314 & \bf0.027 & 18  & \bf25.716 & \bf0.068 &  6    &   \bf0.9          \\ 
C19 & 3:18:26.4   & $-66$:34:26.4  & \bf4.14  & \bf25.126 & \bf0.027 & 20  & \bf25.648 & \bf0.054 &  13   &   \bf1.0          \\ 
C20 & 3:18:22.8   & $-66$:35:16.8  & \bf2.99  & \bf25.416 & \bf0.052 & 18  & \bf25.641 & \bf0.134 &  3    &   \bf1.4          \\ 
C21 & 3:18:24.0   & $-66$:34:40.8  & \bf3.93  & \bf25.571 & \bf0.024 & 20  & \bf25.816 & \bf0.068 &  4    &   \bf0.4          \\ 
C22 & 3:18:26.4   & $-66$:34:19.2  & \bf3.12  & \bf25.630 & \bf0.040 & 20  & \bf26.038 & \bf0.130 & 3     &   \bf1.9         \\ 
C23 & 3:18:30.72  & $-66$:34:51.6  & \bf5.00  & \bf24.965 & \bf0.025 & 20  & \bf25.474 & \bf0.045 &  17   &   \bf2.0        \\ 
C24 & 3:18:23.76  & $-66$:45:32.4  & \bf4.50  & \bf25.191 & \bf0.028 & 19  & \bf25.732 & \bf0.090 & 5     &   \bf1.4        \\ 
C25 & 3:18:28.32  & $-66$:35:09.6  & \bf4.29  & \bf25.360 & \bf0.021 & 19  & \bf25.851 & \bf0.057 & 5     &   \bf0.7         \\ 
C26 & 3:18:27.36  & $-66$:34:48.0  & \bf3.99  & \bf25.551 & \bf0.046 & 19  & \bf26.127 & \bf0.134 & 4     &   \bf1.0        \\ 
\hline
\end{tabular}
\end{table}

\begin{table}
\centering
\caption{\bf Error Budget}
\begin{tabular}{l c c}

    \hline
    \hline
\bf     Source               & \bf Random Error          & \bf Systematic Error \\
    \hline
\bf    Stellar Photometry    & \bf 0.076                  &\bf  ...              \\
\bf    Widths of PL relation & \bf 0.068                  &\bf  ...              \\
    $\bm{\mu_{LMC}}$           &\bf  0.008                  &\bf 0.047            \\
\bf    LMC PL relation       & \bf ...                    & \bf 0.011            \\
\bf    Metallicity Correction & \bf ...                   & \bf 0.035            \\
    \hline 
\end{tabular}
\end{table}

\clearpage

\clearpage
\begin{figure}[htbp]
\includegraphics[clip=true,bb=50 50 560 701]{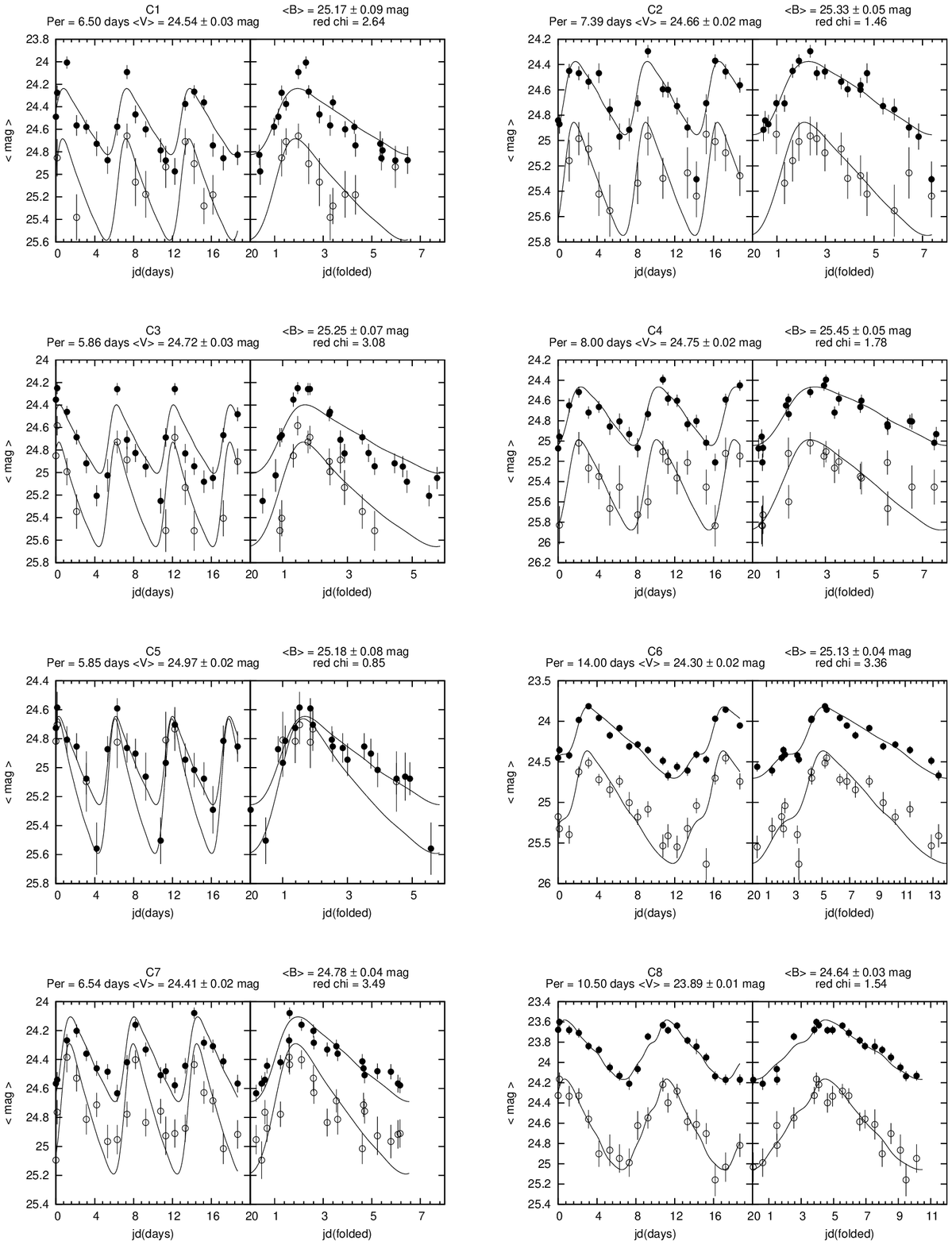}
\caption{\label{LC2}Light curves of the 26 Cepheids}
\end{figure}
 
\begin{figure}[htbp]
\includegraphics[clip=true,bb=50 50 560 701]{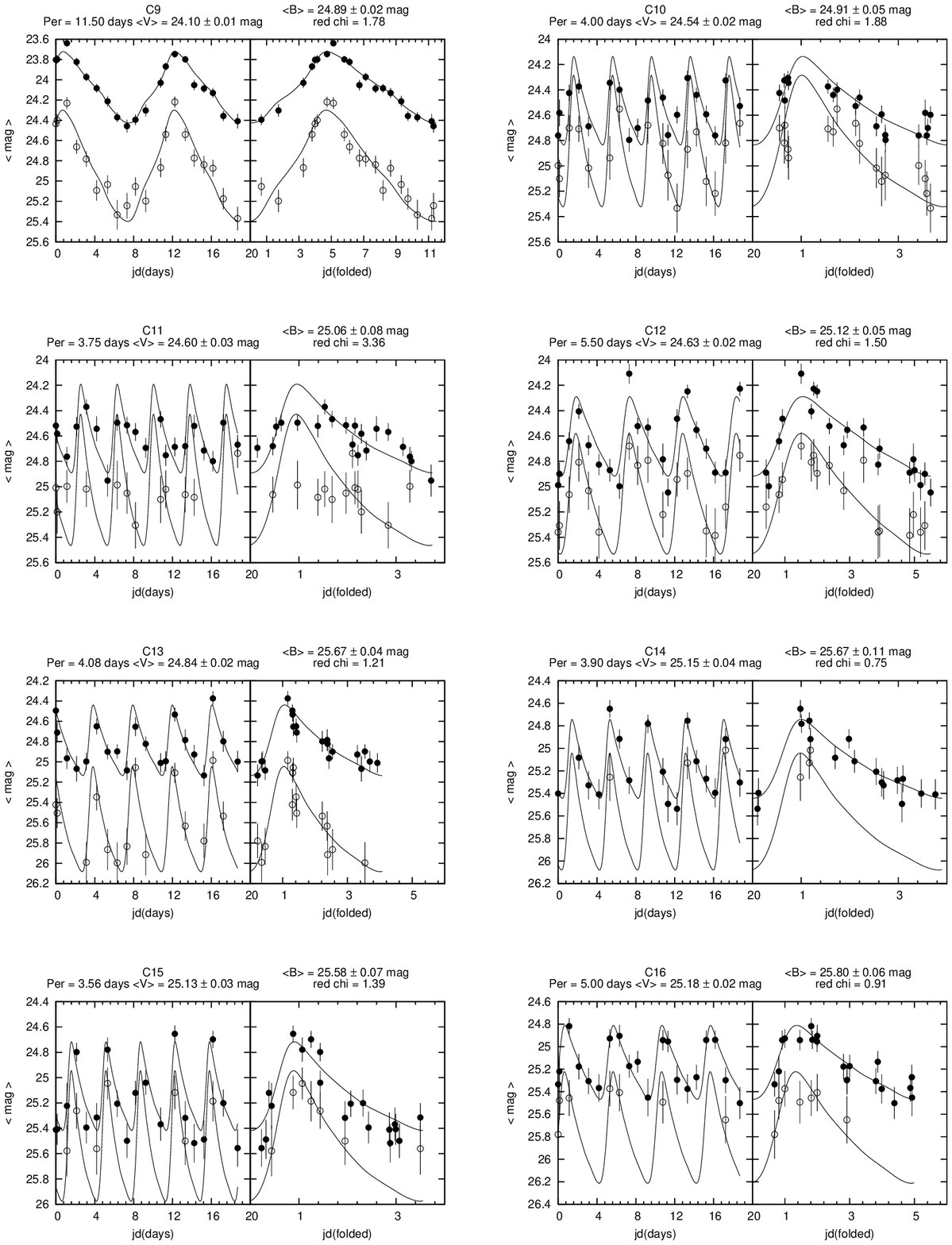}
\contcaption{Light curves of the 26 Cepheids}
\end{figure}
\clearpage
\begin{figure}[htbp]
\includegraphics[clip=true,bb=50 50 560 701]{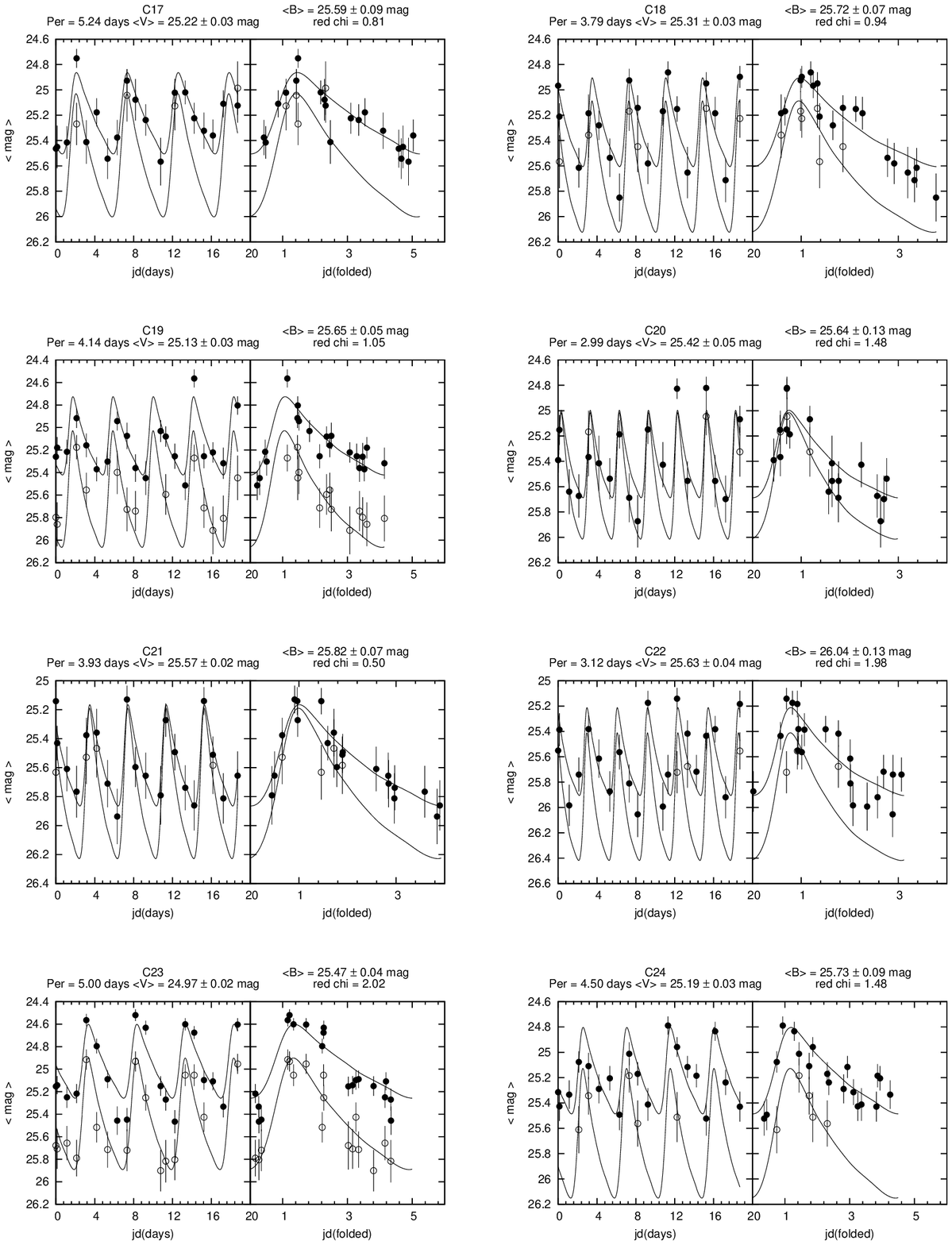}
\contcaption{Light curves of the 26 Cepheids}
\end{figure}
\clearpage
\begin{figure}[htbp]
\includegraphics{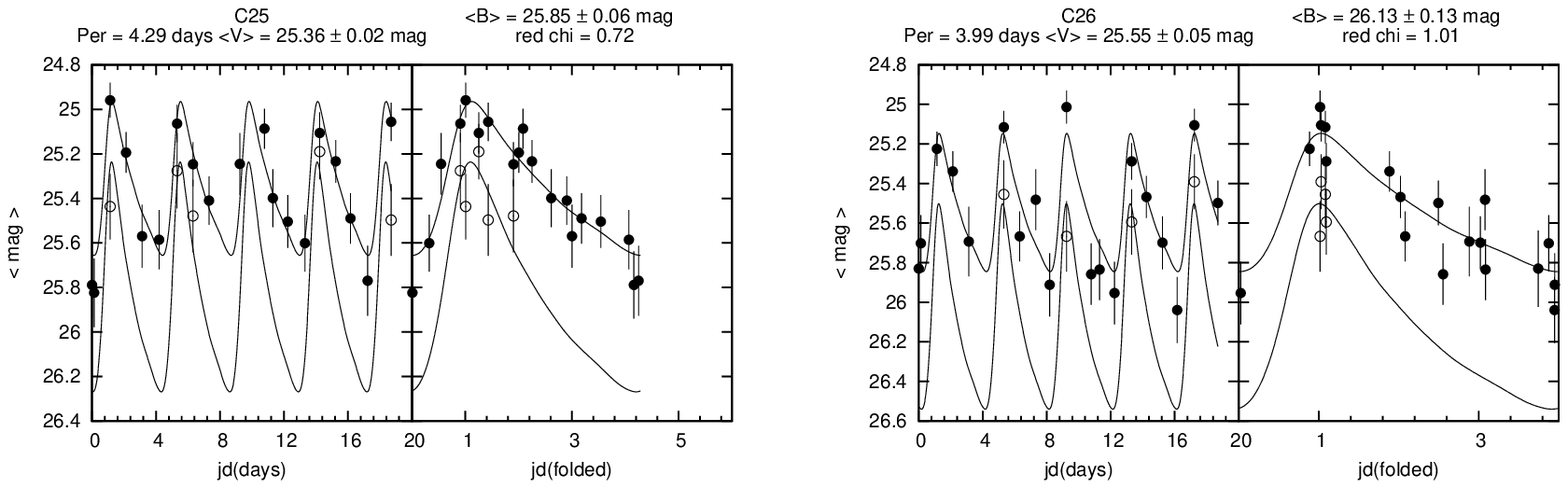}
\contcaption{Light curves of the 26 Cepheids}
\end{figure}

\begin{figure}[htbp]
\includegraphics{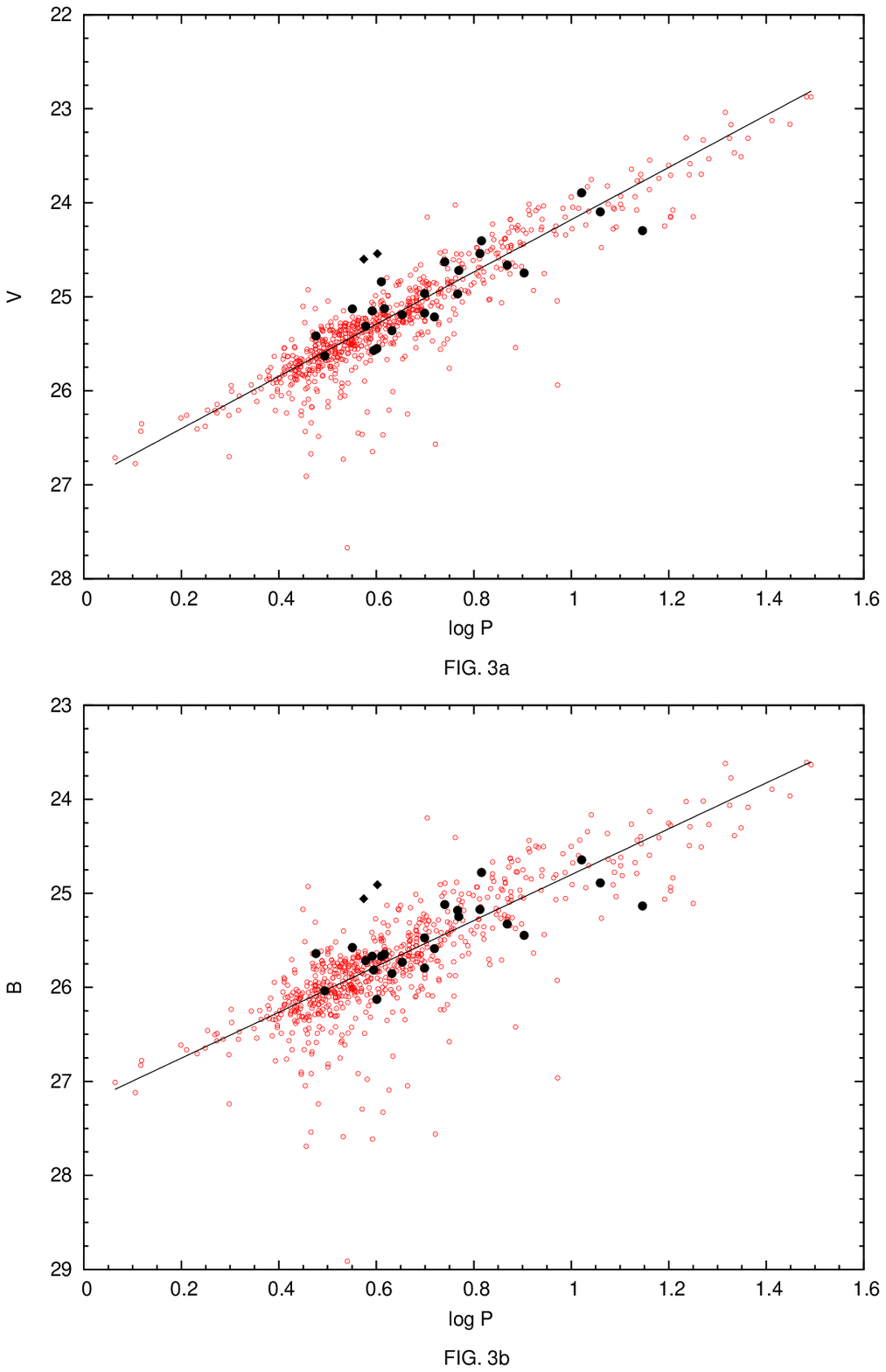}
\caption{\label{PL_LMC} \textbf {(a) V- band PL relation for 26 NGC 1313 Cepheids (black dots), while the black filled diamonds show excluded ones.
It plotted with respect to the LMC Cepheid (red open circles) 
shifted by 9.89 mag in relative apparent modulus.
The magnitudes are plotted vs the logarithm of the period in units of days.
The solid line is the shifted best fit to LMC Cepheid PL relation on V band.
(b) B- band PL relation for 26 NGC 1313 Cepheids (black dots), while the black filled diamonds show excluded ones.
It plotted with respect to the LMC Cepheid (red open circles)
shifted by 9.87 mag in relative apparent modulus.
The solid line is the shifted best fit to LMC Cepheid PL relation on B band.}}
\end{figure}

\clearpage
\begin{figure}[t]
\includegraphics[clip=true,bb=50 50 410 302]{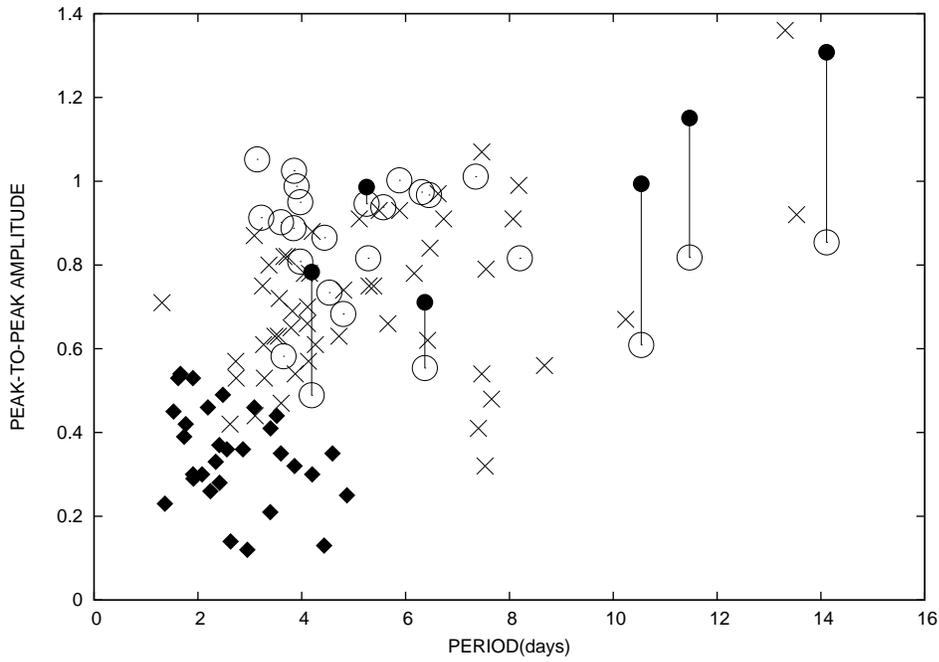}
\caption{\label{Cepheid_classification}Amplitude-period diagram for Cepheids including: 26 $V$-band NGC\,1313 Cepheids  (open circles), 6 $B$-band with detections \textbf{(filled circles)} in all 20 observations. NGC1313 Cepheids (filled dots),  52 LMC FO Cepheids (\textbf{filled} diamonds)
and 32 LMC FU Cepheids (diagonal crosses) from \cite{Beaulieu1995}. The NGC\,1313 Cepheids
clearly belong to the FU type.}
\end{figure}

\end{document}